PAPER

# Estimating Racial and Ethnic Healthcare Quality Disparities Using Exploratory Item Response Theory and Latent Class Item Response Theory Models

S-L. Normand,[1,2,*] K. Zelevinsky[1] and M. Horvitz-Lennon[3]

[1]Department of Health Care Policy, Harvard Medical School, Boston, 02115, MA, USA, [2]Department of Biostatistics, Harvard T. H. Chan School of Public Health, Boston, 02115, MA, USA and [3]The RAND Corporation, Boston, 02116, MA, USA

*Corresponding author. sharon@hcp.med.harvard.edu

## Abstract

Healthcare quality metrics refer to a variety of measures used to characterize what should have been done or not done for a patient or the health consequences of what was or was not done. When estimating healthcare quality, many metrics are measured and combined to provide an overall estimate either at the patient level or at higher levels, such as the provider organization or insurer. Racial and ethnic disparities are defined as the mean difference in quality between minorities and Whites not justified by underlying health conditions or patient preferences. Several statistical features of healthcare quality data have been ignored: quality is a theoretical construct not directly observable; quality metrics are measured on different scales or, if measured on the same scale, have different baseline rates; the construct may be multidimensional; and metrics are correlated within-individuals. Balancing health differences across race and ethnicity groups is challenging due to confounding. We provide an approach addressing these features, utilizing exploratory multidimensional item response theory (IRT) models and latent class IRT models to estimate quality, and optimization-based matching to adjust for confounding among the race and ethnicity groups. Quality metrics measured on 93,000 adults with schizophrenia residing in five U.S. states illustrate approaches.

**Contact:** sharon@hcp.med.harvard.edu

## Introduction

The U.S. Institute of Medicine defines the quality of healthcare as *the degree to which healthcare services for individuals*

*and populations increase the likelihood of desired health outcomes and are consistent with current professional knowledge* (Institute of Medicine, 1990). Numerous professional organizations have promulgated quality measures for a variety of uses, including improving care, selective contracting with healthcare providers, and reimbursement of services delivered (Kronick et al., 2000; Winkelman and Damler, 2008). Healthcare quality measurement is not confined to the U.S. – for instance, the United Kingdom (Minchin et al., 2018), Belgium (Gerkens et al., 2023), France (Pouvourville, 1997), and Portugal (Ramalho et al., 2022) also measure and report quality. In the U.S., the National Committee for Quality Assurance routinely collects administrative billing data to create the healthcare Effectiveness Data and Information Set (HEDIS) indicators on behalf of State Medicaid programs and use these measures to describe the quality of publicly-funded healthcare services (National Committee for Quality Assurance, 2010). Medicaid is a joint state and federal U.S. program that provides health coverage to low-income adults, children, and people with disabilities. The program is large – in 2020, Medicaid included 77.8 million enrollees with total health expenditures of 666.3 billion dollars (Center for Medicaid and CHIP Services (2023)).

The types of quality metrics fall broadly into 3 categories: process measures that characterize whether a patient received recommended care, such as receiving a needed medication; outcome measures that characterize the outcomes of healthcare, such as health status or patient satisfaction; and structural measures that describe the healthcare provider's capacity to deliver healthcare, such as the size of nursing staff.

Healthcare *disparities* refer to the unequal treatment of patients based on race or ethnicity, gender, geography, or other patient characteristics. The U.S. Institute of Medicine (2002) defines a racial and ethnic disparity as a difference in treatment provided to members of different racial and ethnic groups that is not justified by the underlying health conditions or treatment preferences of patients. This definition purposefully excludes adjustment for socioeconomic differences when measuring racial and ethnic disparities. Rather, socioeconomic differences are viewed as potential mediators of racial and ethnic disparities.

In this article, we assess the quality of care for individuals with schizophrenia, a disabling mental illness, associated with chronic medical comorbidities and premature mortality (Weinberger, 2011). These chronic medical comorbidities also complicate treatment, compound disability, and increase healthcare utilization (Parks et al. (2006); Druss et al. (2000)). In the U.S., Medicaid is a key payer for care received by individuals with schizophrenia because most individuals qualify for Medicaid due to the disability caused by the disease(Khaykin et al., 2010; Frank and Glied, 2006). County governments also play a prominent role in the healthcare of people with schizophrenia – state funds directed to counties and other local governments accounted for 24% of all mental health spending in the U.S. during 2015 (SAMHSA, 2019). County variation in treatment patterns for serious mental illness have been reported (Horvitz-Lennon et al., 2009) and may be related to the strength of the county's Medicaid healthcare infrastructure, including availability of and accessibility to specialty care providers, such as psychiatrists (Cummings et al., 2013).

Table 1 summarizes quality information on 93,000 Medicaid beneficiaries aged between 18 and 64 diagnosed with schizophrenia in five U.S. states between 2010 and 2013. Black, Latinx, and White beneficiaries comprise 44%, 15%, and 40%, respectively, of the cohort. Care for individuals is assessed over 12 months. Twenty-three mental health quality metrics are reported and cover pharmacological treatments, acute care services, and ambulatory care services. Each measure is recommended by professional societies or other healthcare entities such as the National Committee for Quality Assurance. All metrics are reported such that a higher percentage corresponds to a more desirable outcome with some exceptions. The exceptions provide denominators for other measures (e.g., items 7, 10). Some metrics are binary, some are ordinal, and some are nominal. While virtually all adults had some ambulatory care, less than 3/4 had quarterly visits. Some variation across race and ethnicity groups is apparent; for instance, quarterly ambulatory visits range from 58% for Black beneficiaries to 66% for White and Latinx beneficiaries.

**Table 1.** Quality measures, overall and by race and ethnicity. Entries are percentages except the last row which reports the mean number (standard deviation) of quality measures met. Ordinal metrics include antipsychotic adherence level (items 4 and 5), follow-up after discharge (items 7+8+9), and follow-up after an emergency department visit (items 10+11+12); nominal metrics include no readmission following a schizophrenia hospitalization (items 17+18) and no readmission following a mental health hospitalization (items 19+20).

| | **All** | **Black** | **Latinx** | **White** |
|---|---|---|---|---|
| Number of Persons | 93,311 | 41,297 | 14,301 | 37,713 |
| 1. Any antipsychotic drug use | 90 | 90 | 93 | 90 |
| 2. Any log-acting injectable antipsychotic drug | 15 | 17 | 14 | 13 |
| 3. Any clozapine use | 4.6 | 2.5 | 4.7 | 7.0 |
| 4. Antipsychotic adherence† $\geq$ 80% | 49 | 39 | 51 | 59 |
| 5. Antipsychotic adherence† $\geq$ 50% | 72 | 64 | 76 | 80 |
| 6. No antipsychotic polypharmacy | 79 | 83 | 78 | 76 |
| 7. Any schizophrenia inpatient discharge | 19 | 20 | 21 | 18 |
| 8. Follow-up within 7 days | 24 | 23 | 26 | 24 |
| 9. Follow-up care 30 days | 36 | 34 | 38 | 36 |
| 10. Any schizophrenia ED visit | 3.6 | 4.4 | 2.3 | 3.1 |
| 11. Follow-up care within 7 days | 37 | 35 | 38 | 39 |
| 12. Follow-up care within 30 days | 47 | 44 | 51 | 51 |
| 13. Any ambulatory mental healthcare visit | 99 | 98 | 98 | 99 |
| 14. Quarterly ambulatory MH care visits | 63 | 58 | 66 | 66 |
| No excessive acute care utilization for: | | | | |
| 15. schizophrenia | 95 | 94 | 95 | 95 |
| 16. mental health | 91 | 91 | 91 | 91 |
| 17. Any schizophrenia hospitalization | 20 | 20 | 22 | 19 |
| 18. No 30-day readmission | 77 | 77 | 76 | 77 |
| 19. Any MH hospitalization | 23 | 23 | 24 | 22 |
| 20. No 30-day readmission | 74 | 74 | 73 | 73 |
| 21. Any MH psychosocial services | 84 | 85 | 83 | 83 |
| 22. Quarterly psychosocial services | 41 | 40 | 37 | 43 |
| 23. Any individual, group, or family psychotherapy | 28 | 36 | 12 | 26 |
| Mean number met (SD) | 10.0 (2.14) | 9.89 (2.21) | 10.0 (2.02) | 10.2 (2.08) |

†Measured as a percentage of days covered. ED = emergency department. MH = mental health

The most common approach to estimate healthcare quality when multiple indicators are measured uses a compensatory scoring algorithm, either averaging or summing the numerical values assigned to the metrics for each individual, typically weighing all quality metrics equally (Higashi et al., 2007). Although less common, conjunctive scoring of the observed responses, which requires a passing performance on each quality metric, has also been used (Nolan and Berwick, 2006). To identify variables associated with quality, investigators have also fitted separate regression models to each quality indicator (Song et al., 2014), examining regression coefficients. Findings focused on the quality of care received by mentally ill individuals have typically reported results separately for each measure (Busch et al., 2008; Cully et al., 2008). We are aware of only two studies that provided a summary measure capturing multiple domains of care (Kilbourne et al., 2010; Hepner et al., 2007) but neither used a psychometric-based approach. Few approaches adopt a measurement framework for quality estimation, that is, viewing quality as a theoretical construct that is not directly observable. In a measurement framework, each quality metric provides observable information about the theoretical quality construct(s) and that construct needs to be inferred. The absence of a measurement approach may introduce bias into summary estimates of quality and quality disparities. Given the policy implications of erroneously describing the quality of care received by some patient subgroups as poor, the adoption of models commensurate with the data-generating process and the theoretical underpinnings of quality is critical.

Approaches to measuring constructs exist – there is a large psychometric literature for measure development. The text by Skondral (2004) and references therein focus on statistical issues as does the text by Baker (1994). Of particular relevance are approaches that utilize item response theory modeling (IRT) in which the metrics (or items) are discrete (Embretson and Reise, 2000). For measuring healthcare quality specifically, Landrum et al. (2000), Normand et al. (2007), Teixeira-Pinto et al. (2008), and He et al. (2010) employ IRT models to assess hospital quality using aggregated patient data while Agniel et al. (2020) use IRT models on aggregated ambulatory service measures to evaluate hospital systems. Landrum et al. (2003) modeled clustered mixed (continuous and discrete) patient-level quality data using latent variables to connect outcomes; Daniels and Normand (2006) proposed methods for clustered longitudinal mixed patient-level quality data. However, both approaches relied on parametric modeling to risk-adjust for underlying health differences.

The goal of this paper is to derive an approach for the estimation of race and ethnicity healthcare quality disparities for categorical data, possibly having a multidimensional structure and requiring risk adjustment.

# Approaches

Let $Y_{ij}$ denote the observed outcome associated with quality metric (or item) $i$ for individual $j$. Response options for the items can be binary or ordinal such that larger values correspond to better quality, or nominal where there is no natural ordering. We assume there are $I$ items and the total number of individuals is $N$. Not all metrics will apply to an individual; for instance, follow-up within 30 days of a hospital admission is only applicable to those discharged alive from a hospital. We let $n_j$ denote the number of measures for which individual $j$ is eligible.

## Common Approaches

The most frequently used estimator of quality received by individual $j$ is

$$Z_j = \frac{\bar{Y}_{\cdot j} - \bar{Y}_{\cdot\cdot}}{s} \text{ where } \bar{Y}_{\cdot j} = \frac{1}{n_j}\sum_{i=1}^{n_j} Y_{ij}, \tag{1}$$

$$\bar{Y}_{\cdot\cdot} = \frac{1}{N}\sum_{j=1}^{N} \bar{Y}_{\cdot j}, \text{ and } s = \sqrt{\frac{\sum_{j=1}^{N}(\bar{Y}_{\cdot j} - \bar{Y}_{\cdot\cdot})^2}{N}}.$$

When individuals are eligible for different numbers of metrics, $\bar{Y}_{\cdot j}$ has been referred to as an *opportunity score* (Anderson et al., 2016) because it reflects the number of opportunities available to a healthcare provider to deliver guideline-concordant or recommended care. Using these approaches, the minority-White quality disparity estimator is

$$\frac{1}{\sum_j x_j}\sum_{j=1}^{N} Z_j x_j - \frac{1}{N - \sum_{j=1}^{N} x_j}\sum_j Z_j(1 - x_j) \tag{2}$$

where $x_j = 1$ if the individual is a minority individual and 0 if the individual is White individual. To adjust for differences in health status across groups, the disparity is estimated using a linear regression model, e.g.,

$$Z_j = \mathbf{w}_j \boldsymbol{\gamma}_r + \mathbf{v}_j \boldsymbol{\gamma}_h + \epsilon_j \tag{3}$$

where $\mathbf{w}_j$ is a $1 \times q$ vector of race and ethnicity indicators, $\boldsymbol{\gamma}_r$ is a $q \times 1$ coefficient vector that characterizes quality disparities, $\mathbf{v}_j$ is a $1 \times p$ vector of health status variables, $\boldsymbol{\gamma}_h$ is a $p \times 1$ coefficient vector meant to balance health status differences among race and ethnicity groups, and $\epsilon_j \sim N(0, \sigma^2)$.

While this approach is simple to interpret and easy to calculate, there are several problems. The approach assumes all metrics use the same response scale which is often unlikely. The continuous or categorical responses could be converted to binary responses. Even when the metrics have the same response options, the underlying frequency of each metric may differ. The optimal (in terms of variance) algorithm for pooling information depends upon the type of measurement error present, which in turn depends on the underlying frequency of the metric. Moreover, researchers assume that the Var$(Z_j)$ is constant across individuals and that the items used to construct $Z_j$ are independent within an individual. Both assumptions are incorrect and conflict with the conceptual model of healthcare quality. Pooling algorithms are not clearly defined if some items are missing. Although multiple imputation could be used to fill in missing values at the metric level, typically single-based imputation, such as using the mean response, is used. If separate regression models are estimated, the same methodological issues remain, global tests of overall race and ethnicity effects are not possible, and missing data complicates inferences (Daskalakis et al. (2002); Yoon et al. (2011)). Finally, adjusting for baseline differences in health status using a parametric model may fail to appropriately balance observed differences. A unified multivariate model avoids many of these issues.

Our point of departure treats the observed quality metrics as manifestations of a single or a few latent variables, and conditional on the latent variables, assumes that the observed outcomes are independent. The latent variables characterize the relationship among the observed quality metrics within an individual. Using algorithms from the causal inference literature, a model-free matching approach balances observed health status differences across groups and a structural model estimates racial and ethnic disparities in the matched sample. Throughout we use trait, dimension, and construct interchangeably.

## Item Response Theory Model

An item response theory model connects categorical responses to item-specific parameters and a latent trait (or traits). Assuming a dichotomous item and a single latent variable, a commonly used model is the 2-parameter logistic (2PL) model defined as

$$P_i(\theta_j) = P(Y_{ij} = 1 \mid \theta_j) = \frac{1}{1 + \exp(-a_i(\theta_j - b_i))} \tag{4}$$

where $\theta_j$ is the underlying latent trait (quality in our setting) for individual $j$, $a_i$ is an item discrimination parameter, and $b_i$ is an item difficulty parameter. While we use a logit link, a probit link would work as well. Assumptions about $\theta_j$ are discussed later. The larger the value of $a_i$, the more discriminating item $i$ is as it reflects the rate of change in the probability of receiving guideline-concordant care to changes in the latent trait. The item difficulty parameter, $b_i$, is the point on the latent trait continuum in which an individual has a 0.50 probability of receiving recommended care. In our setting, the difficulty parameter measures how difficult it is to provide or to receive guideline-concordant care. If $a_1 = a_2 = \cdots a_I = 1$, then the IRT model has a random intercept only (referred to as a *Rasch* model), and $\bar{Y}_{\cdot j}$ is a sufficient statistic for $\theta_j$ conditional on $b_i$. Only when the item discrimination parameters are equal is an unweighted average (or sum) of the binary metrics (e.g., Equation 1) suitable for estimating the latent construct. If some of the discrimination parameters differ across items, then the *pattern* of outcomes, rather than the total or average outcome, is important.

Without loss of generality, we reparameterize the 2-parameter logistic IRT model as

$$P_i(\theta_j) = \frac{\exp(a_i\theta_j + d_i)}{1 + \exp(a_i\theta_j + d_i)}. \tag{5}$$

In this parameterization, $d_i$ is referred to as item easiness so that $b_i = -d_i/a_i$. For quality measures having polytomous responses, a nominal IRT model is given by

$$P_{ik}(\theta_j) = P(Y_{ij} = k \mid \theta_j) = \frac{\exp(a_i^k\theta_j + d_i^k)}{\sum_{p=0}^{K-1}\exp(a_i^p\theta_j + d_i^p)} \tag{6}$$
$$\text{with } a_i^0 = d_i^0 = 0$$

where quality metric $i$ has response options $k = 0, 1, \cdots, K_i - 1$. Here, the discrimination and easiness parameters vary not only by quality metric but also by the response option associated with the metric. The point on the latent trait scale where an individual is equally likely to have response options to adjacent categories is

$$\frac{d_i^{k-1} - d_i^k}{a_i^k - a_i^{k-1}}.$$

For ordinal measures, we consider the graded IRT model,

$$\begin{aligned}
P(Y_{ij} \geq 0 \mid \theta_j) &= 1 \\
P(Y_{ij} \geq k \mid \theta_j) &= \frac{\exp(a_i\theta_j + d_i^k)}{1+\exp(a_i\theta_j + d_i^k)};\ k = 1, \cdots, K_i - 1 \\
P(Y_{ij} \geq K_i \mid \theta_j) &= 0 \text{ so that} \\
P(Y_{ij} = k \mid \theta_j) &= P(Y_{ij} \geq k \mid \theta_j) - P(Y_{ij} \geq k+1 \mid \theta_j).
\end{aligned}$$

Examination of the item easiness parameters provides some insight into the appropriateness of the ordinal assumption. Other IRT models are available that vary by the type of restrictions placed on the $d_i^k$ but we do not consider these in this manuscript.

## *Assumptions*

Several assumptions are made when using IRT models.

*Assumption 1 (local independence).* Responses within individuals are independent conditional on the latent variable, $\theta_j$

$$P(\mathbf{Y}_j = \mathbf{y}_j \mid \mathbf{a}, \mathbf{d}, \theta_j) = \prod_{i=1}^{n_j} P(Y_{ij} = y_{ij} \mid a_i, d_i, \theta_j)$$

where $y_{ij}$ is the observed response, $\mathbf{y}_j = \{y_{1j}, y_{2j}, \cdots, y_{n_j j}\}$, $\mathbf{a} = \{a_1, a_2, \cdots, a_I\}$, and $\mathbf{d} = \{d_1, d_2, \cdots, d_I\}$. This assumption is common in random effects models and could be relaxed by permitting conditional associations among sets of items.

*Assumption 2 (monotonicity).* Increasing $\theta_j$ implies that the probability of receiving care consistent with the guideline does not decrease. With $a_i > 0$, Assumption 2 implies that the probability distribution of $Y_{ij}$ moves toward higher values as $\theta_j$ increases.

*Assumption 3 (latent trait distribution).* Some assumptions about the distribution of latent traits must be made to provide a metric for the trait scale. The most common assumption is that of Normal latent traits.

*3a: Normal Latent Trait:* $\theta_j \overset{iid}{\sim} N(\mu, \Sigma)$, $j = 1, \cdots, N$. This is likely the most familiar latent trait model, and under this assumption, the 2-PL and graded IRT model result. Other assumptions about the latent trait can be adopted, such as mixtures of Normal distributions, log-linear, or kernel smoothing. We also assume a discrete distribution.

*3b: Latent Class* (Goodman (1974)): $\theta_j, j = 1, 2, \cdots, N$ are iid with $C$ support points $\xi_1, \xi_2, \cdots, \xi_C$ with corresponding probabilities $\pi_1, \pi_2, \cdots, \pi_C$ having

$$\pi_{cj} = p(\theta_j = \xi_c), \text{ such that } \sum_{c=1}^{C} \pi_{cj} = 1.$$

The $\xi_c$s are locations that depend on the class. In the latent class model, individuals belong to one of $C$ classes and class membership is unknown. Within each class, individuals are homogeneous with respect to the underlying construct. The model is semi-parametric and thus offers some robustness protection. Classes are categories of a latent variable so that the *prior* probability that beneficiary $j$ belongs to class $c$, $\pi_{cj}$, is a model parameter.

## *Multidimensionality*

A unidimensional model assumes that the variation in each quality metric is only affected by the variation in a single construct and not in any other related construct. In our setting, quality can be multidimensional given that metrics cover a wide range of healthcare services: pharmacological treatments, acute care services, and ambulatory care services. There are various modeling options, each depending on what is hypothesized about quality. Suppose that there are $s = 1, 2, \cdots, S$ constructs. A multidimensional 2-PL (M2PL) is given by

$$P_i(\boldsymbol{\theta}_j) = \frac{\exp(\sum_{s=1}^{S} a_{is}\theta_j^s + d_i)}{1+\exp(\sum_{s=1}^{S} a_{is}\theta_j^s + d_i)} \tag{7}$$

where $\boldsymbol{\theta}_j = (\theta_j^1, \theta_j^2, \cdots, \theta_j^S)^\top$ is an $S \times 1$ vector of quality constructs, $\mathbf{a}_i = (a_{i1}, a_{i2}, \cdots, a_{iS})^\top$ is an $S \times 1$ vector of discrimination parameters, and $d_i$ is an easiness parameter for item $i$. This model assumes that the variance in each quality metric is a weighted function of up to $S$ common traits so that *quality* is what the primary traits have in common. In the multidimensional setting, the **a** parameters indicate how well an item discriminates but this depends on the direction that is being measured in the $\theta$-space. Model (7) is referred to as a compensatory multidimensional IRT model because it permits individuals' higher placement on one trait to overcome low positions on another trait in the probability of receiving guideline-concordant care associated with a metric. Partially compensatory multidimensional IRT models that do not have this property exist but will not be discussed further. The interested reader should consult Reckase (2009).

A multidimensional latent class IRT model assumes that each item is related to only one latent trait in the set of latent traits. This assumption is not imposed by the multidimensional IRT models described above. Let $\mathcal{I}_s$ denote disjoint subsets of items, $i = 1, \cdots, I$ describing the items relating to trait $s$ and let

$$\delta_{is} = \begin{cases} 1 & \text{if } i \in \mathcal{I}_s \\ 0 & \text{otherwise.} \end{cases}$$

Then for binary items, for instance, the multidimensional latent class IRT model takes the form

$$P_i(\boldsymbol{\theta}_j) = \frac{1}{1 + \exp(-a_i(\sum_{s=1}^{S} \delta_{is}\theta_j^s - b_i))}. \quad (8)$$

ASSUMPTIONS

The assumptions in the multidimensional setting for the models considered in this paper are similar to those made in the unidimensional setting.

*Assumption M1: (local independence).* Responses within individuals are independent conditional on $\boldsymbol{\theta}_j$.

*Assumption M2: (monotonicity).* The probability of the response occurring increases as any element in $\boldsymbol{\theta}_j$ increases.

*Assumption M3: (latent trait distribution).* For Normal latent traits, $\boldsymbol{\theta}_j \sim N_S(\boldsymbol{\mu}, \Sigma)$ for $j = 1, \cdots, N$. For the discrete setting, $\boldsymbol{\theta}_j$ has a discrete distribution with support points $\boldsymbol{\xi}_1, \boldsymbol{\xi}_2, \cdots, \boldsymbol{\xi}_C$ and corresponding probabilities $\pi_1, \pi_2, \cdots, \pi_C$ such that

$$\pi_{cj} = p(\boldsymbol{\theta}_j = \boldsymbol{\xi}_c), \text{ such that } \sum_{c=1}^{C} \pi_{cj} = 1.$$

*Including Covariates*

QUALITY DISPARITIES

Rather than assuming that all the latent variables have the same distribution, covariates may be introduced. In our setting, summaries of $\theta_j$ by race and ethnicity groups are examined to detect quality disparities. More formal associations are estimated using a structural regression model. In the Normal latent trait model, the latent trait is linked to covariates using

$$\theta_j = \mathbf{w}_j \boldsymbol{\gamma} + \epsilon_j \quad (9)$$

where $\mathbf{w}_j$ is a $1 \times q$ vector of dummy variables indicating if the individual is from a Black or Latinx race and ethnicity group, and $\epsilon_j$ a vector of errors from a $N(0, 1)$ distribution. Negative values of vector components $\boldsymbol{\gamma}$ correspond to a lower quality for minorities compared to White individuals. If $\boldsymbol{\theta}_j$ is multidimensional, then a multivariate model can be estimated, allowing the relationship of $\mathbf{w}_j$ with each trait $\theta_j^s$ to differ.

In contrast, in the latent class model, the class probabilities, $\pi_{cj} = p(\theta_j = \xi_c \mid \mathbf{w}_j)$, are modeled. For instance, a multinomial logit model assumes

$$\begin{aligned} \pi_{cj} &= \frac{\exp(\mathbf{w}_j \boldsymbol{\gamma}_c)}{1 + \sum_{c^*=2}^{C} \exp(\mathbf{w}_j \boldsymbol{\gamma}_{c^*})} \quad c = 2, \cdots, C \\ \pi_{1j} &= \frac{1}{1 + \sum_{c^*=2}^{C} \exp(\mathbf{w}_j \boldsymbol{\gamma}_{c^*})} \end{aligned} \quad (10)$$

which links covariates to the probability of class $c$ membership. Alternatively, a cumulative logit model imposes ordering constraints on intercepts within $\boldsymbol{\gamma}_c$ while forcing the other components of $\boldsymbol{\gamma}_c$ to be the same across all classes (Bartolucci, Dardanoni, and Peracchi (2016)). Without loss of generality, order $\xi_1, \xi_2, \cdots, \xi_C,$, and

$$\frac{\sum_{c>c} \pi_{cj}}{\sum_{c\leq c} \pi_{cj}}, \quad c = 1, 2, \cdots, C-1$$
$$\text{where } \log\left(\frac{\sum_{c>c} \pi_{cj}}{\sum_{c\leq c} \pi_{cj}}\right) = \phi_c + \mathbf{w}_j^* \boldsymbol{\gamma}^* \tag{11}$$

such that $\phi_1 < \phi_2 < \cdots < \phi_{C-1}$, with $\mathbf{w}_j^*$ and $\boldsymbol{\gamma}^*$ having dimension $1 \times q - 1$. If $\boldsymbol{\theta}_j$ is multidimensional, the multinomial logit model is adopted.

### Adjusting for Health Status

Health differences are likely between race and ethnicity groups and balancing these differences can be challenging. To avoid misspecification of the regression model and to strengthen the balance of patient health status between race and ethnicity groups, approaches other than regression could be adopted. In template matching (Bennett et al., 2020), a simple random sample of individuals is taken to create a *template* that reflects the target population of interest. Within each race and ethnicity group, $r$, individuals are matched to the characteristics in the template using a constrained optimization procedure. The goal is to

$$\underset{c,z}{\text{minimize}} \sum_v \sum_p c_{v,p} \text{ such that } \left|\sum_{s=1}^{N_r} z_s - N_{v,p}\right| \leq c_{v,p}$$
$$\text{and } \sum_{s=1}^{N_r} z_s = T \quad \forall v \in \mathcal{V}$$

where $c_{v,p}$ is a measure of imbalance for category $p$ of covariate $v$; $z_s \in \{0, 1\}$ is a binary indicator assuming a value of 1 if the $s$th observation in a racial and ethnic group $r$ having sample size $N_r$ is matched to the template sample and 0 otherwise; $N_{v,p}$ is the number of observations in the template sample having category $p$ for the $v$th covariate; $T$ is the size of the template, and all covariates are discretized. This approach is attractive because it is model-free and specific constraints can be selected. Using the resulting matched sample, the simpler structural model described in Equations (9) and (10) provides estimates of racial and ethnic disparities. Regression estimates based on the matched template reduce the dependence of the validity of the estimated disparities on the correct specification of the model.

### *Identifiability*

Identifiability of the model is generally approached in one of two ways: (1) The diagonal elements of $\Sigma$ are set to 1 and $\mu$ is equated to 0. If a structural regression model is estimated, this implies $\mu = \sum_j \mathbf{w}_j \boldsymbol{\gamma} = 0$. Otherwise, (2) for each latent trait, one discriminating parameter is set to 1 and one difficulty parameter is set to 0. Other constraints are also necessary; for unidimensional models, at least 3 items are needed. For multidimensional traits, an orthogonal rotation is used to address rotational indeterminancies. For a comprehensive discussion of identification in latent variable models from an analytical perspective, see Skondral (2004)[Chapter 5]. Reckase (2009) provides guidance on rotations to facilitate interpretation of multidimensional Normal latent trait structure once estimates have been obtained.

### *Inference*

Assuming the traits are independent and identically distributed (possibly within subsets defined by covariates $\mathbf{w}$) from a population with a Normal distribution, the marginal likelihood function for the data is

$$L(\Theta, \mathbf{a}, \mathbf{d}, \Sigma, \boldsymbol{\gamma} \mid \mathbf{y}) =$$
$$\prod_{j=1}^{N} \left\{ \int_\theta \prod_{i=1}^{n_j} P(Y_{ij} = y_{ij} \mid \mathbf{a}_i, \mathbf{d}_i, \theta_j) N(\theta_j \mid \boldsymbol{\gamma}, \Sigma) d\theta \right\} \tag{12}$$

where $\Theta = \{\theta_1, \theta_2, \cdots, \theta_N\}$. Estimation of model parameters uses the expectation-maximization (EM) algorithm via Gaussian-Hermite quadrature for each unique response vector. In models with more than 3 traits, Quasi Monte Carlo EM (QMCEM) estimation methods are implemented (Jank, 2005). The QMCEM algorithm is a stochastic version of the EM algorithm that replaces the E-Step by a Monte Carlo approximation and uses a sequence of points rather than purely random sampling.

For the latent class model, the marginal likelihood is a mixture distribution

$$L(\Theta, \boldsymbol{\xi}, \mathbf{a}, \mathbf{d}, \boldsymbol{\pi}, \boldsymbol{\gamma} \mid \mathbf{y}) = \prod_{j=1}^{N} \left\{ \sum_{c=1}^{C} \pi_{cj}(\boldsymbol{\gamma}) \prod_{i=1}^{n_j} P(Y_{ij} = y_{ij} \mid \theta_j = \xi_c, \mathbf{a}, \mathbf{d}) \right\} \quad (13)$$

where $\pi_c(\gamma)$ denotes that the probability depends on a regression function. As with the Normal latent trait model, the EM algorithm is used for estimation.

For multidimensional models, a multivariate Normal distribution is used in Equation (12). For the latent class model, Equation (13) becomes

$$\prod_{j=1}^{N} \left\{ \sum_{c=1}^{C} \pi_{cj}(\boldsymbol{\gamma}) \prod_{s=1}^{S} \prod_{i \in \mathcal{I}_s} P(Y_{ij} = y_{ij} \mid \theta_j = \xi_c, \mathbf{a}, \mathbf{d}) \right\}. \quad (14)$$

To avoid multimodality of the likelihood in latent class models, the EM algorithm is initialized by a deterministic rule and by a multi-start strategy based on random starting values: the class weights are drawn from a continuous uniform distribution between 0 and 1, normalized to sum up to 1; the other parameters are generated from independent standard normal distributions.

### *Distribution of the Latent Traits or Class Probabilities*

The distribution of latent traits provides the key information on quality of healthcare care for our application. For Normal latent traits, the mean

$$E(\theta \mid \mathbf{y}) = \int \theta P(\theta \mid \mathbf{y}) d\theta \quad (15)$$

for each individual provides information regarding where, on the quality scale, an individual resides. Higher mean values support receipt of guideline-concordant care. For the latent class models, we can assign individuals to a single latent class $h*$ using

$$h* = \max_{c=1,\cdots C} p(\theta_j = \xi_c \mid \mathbf{y}_j) \quad (16)$$

and obtain a distribution of quality for each enrollee in the sample. Within each class, the race and ethnicity distribution can be summarized.

### *Model Fit*

#### NORMAL LATENT TRAIT IRT MODEL

Dimensionality and general model fit are assessed using the Aikaike Information Criteria (AIC) and the Bayesian Information Criteria (BIC) statistics by estimating models of increased complexity, and selecting models with the lowest statistic. Overall model fit is also assessed using the root mean squared error of approximation (RMSEA) proposed by Maydeu-Olivares and Joe (2014) with values of $< 0.05$ desirable. The normality of the latent traits is assessed through quantile-quantile plots. Local independence is examined by correlating the residuals (deviation of the observed and model-predicted item scores) for pairs of items with values near zero desirable.

#### LATENT CLASS IRT MODEL

In the latent class models, the number of traits and item allocation to each trait are required. Factor analysis or hierarchical clustering (Bartolucci (2007)) can be used to determine these. The number of classes is determined using AIC and BIC; with a fixed number of classes, the number of dimensions can be tested using BIC or the likelihood ratio test (Bartolucci, Bacci, and Gnaldi (2014)). Local independence may be assessed by fitting extended IRT models that relax the monotonicity assumption. For instance, a model that included a set of association parameters for specific pairs of responses after conditioning on the latent trait could be tested against the standard model that assumes standard local independence.

### *Validation*

If an outcome or item is available, the model is validated by correlating the predicted item scores or predicted class assignments with the outcome. If the relationships are in the expected directions, then model validation is supported.

## Application

Using healthcare billing data, we identified all Medicaid beneficiaries aged 18 - 64 years of age residing in one of five U.S. states (California, Georgia, Michigan, Mississippi, New Jersey) having (1) at least two outpatient visits from two different dates during 12 months with a primary or secondary diagnosis of schizophrenia, or at least one inpatient discharge with a primary schizophrenia diagnosis; (2) continuous enrollment 6 months before their qualifying schizophrenia diagnosis; and (3) continuous enrollment in Medicaid for 9 out of 12 months of enrollment, with no more than a month's gap after the qualifying diagnosis. These states were selected to ensure the inclusion of sufficient Black and Latinx enrollees. We refer to the first date of the qualifying schizophrenia diagnosis as the *index* date and quality of care was assessed in the 12 months following the index date.

Columns 1 through 4 in Table 2 summarize the cohort, overall, and by race and ethnicity group. Approximately half of the cohort lived in California, had a mean age of 42 years, a quarter had other serious psychiatric comorbidities, 1 in 10 had substance use disorder comorbidity, and close to half had cardiometabolic morbidity. Most individuals received supplemental security income due to a physical or mental health disability which we use to characterize their health status along with other comorbidities. As expected, race and ethnicity distributions varied by state, with many Latinx enrollees residing in California. Some differences in comorbidities between the races and ethnic groups are evident, such as cardiometabolic comorbidity and cognitive comorbidity.

We combined the response options for the measures in Table 1 that occur in less than 2% of the cohort with the closest response option and, if not possible, eliminated the metric from further consideration. Items that are conditional on another item, e.g., responses on one affect the responses on others, were combined to help with the monotonicity assumption. While some indicators have ordinal responses, nominal response options were also used. We excluded quarterly receipt of psychosocial services (item 22) from all models to use as a validation measure. To adjust for health status, we created a template sample by randomly sampling 24,000 enrollees to match 8,000 enrollees per race and ethnicity group to the template. This number of observations should be sufficient to estimate multidimensional IRT models. Using the `cardmatch` function in the `designmatch` `R` package, we matched individuals from each race and ethnicity group to the template by constraining the means of the covariates to be equal across race and ethnicity groups. The balance of observed confounders was examined by comparing the distribution of the covariates across race and ethnicity covariates. We used the covariates presented in Table 2 except for the state of residence because the state may be on the causal pathway, as it reflects the availability of social and healthcare resources. Year was included to capture cohort effects and age was categorized into quartiles.

### Exploratory Normal Latent Trait IRT Model

We initially estimated 2-parameter logistic, ordinal, and nominal response IRT models according to the original item response type, assuming between 1 and 3 traits using the `mirt` function in the `mirt` `R` package (Chalmers, 2012), with $\boldsymbol{\theta} \sim N_S(\boldsymbol{\mu}, \Sigma)$, setting the diagonal elements of $\Sigma$ to 1 and $\boldsymbol{\mu} = \mathbf{0}$. We determined the nature of each trait by examining the relative loadings across items. We used an orthogonal rotation, varimax, constrained the $\mathbf{a}$ parameters to be positive, and considered loadings greater than $|0.20|$ meaningful. Local independence was assessed by examining pairwise item residual correlations. Items were eliminated if the fit was poor and models re-estimated.

### Exploratory Latent Class IRT Model

Latent class models were estimated using the `MultiLCIRT` package in `R` (Bartolucci, Bacci, and Gnaldi, 2014); the EM algorithm was initialized using the recommended strategies of a deterministic rule and again using random starting values.

**Table 2.** Cohort characteristics, by race and ethnicity group.

| Characteristic | (1) All | (2) Black | (3) Latinx | (4) White | (5) Template |
|---|---|---|---|---|---|
| Number (%) | 93,311 (100) | 41,297 (44) | 14,301 (15) | 37,713 (40) | 24,000 (26) |
| Female | 46 | 47 | 41 | 47 | 46 |
| Mean age, (SD) | 42 (13) | 42 (13) | 37 (13) | 43 (13) | 42 (13) |
| Index Year | | | | | |
| 2010 | 52 | 50 | 52 | 54 | 53 |
| 2011 | 20 | 21 | 19 | 20 | 20 |
| 2012 | 15 | 16 | 15 | 13 | 14 |
| 2013 | 13 | 14 | 14 | 13 | 13 |
| State of Residence | | | | | |
| California | 54 | 38 | 91 | 58 | 62 |
| Georgia | 13 | 21 | 1.0 | 9.5 | 11 |
| Michigan | 18 | 22 | 2.1 | 21 | 14 |
| Mississippi | 5.7 | 9.8 | 0.15 | 3.5 | 4.6 |
| New Jersey | 8.8 | 9.7 | 6.0 | 8.9 | 8.4 |
| SSI | 91 | 92 | 89 | 92 | 92 |
| †Comorbidity | | | | | |
| Cognitive | 3.9 | 3.7 | 2.5 | 4.7 | 3.9 |
| ‡Psychiatric Disorders | 25 | 24 | 26 | 37 | 25 |
| Substance Use Disorder | 11 | 11 | 10 | 11 | 11 |
| Diabetes | 10 | 11 | 9.9 | 9.0 | 10 |
| Cardiometabolic | 44 | 45 | 38 | 44 | 43 |
| Other Medical | 21 | 20 | 17 | 24 | 21 |
| †Schizophrenia Inpatient Days | | | | | |
| 0 | 85 | 85 | 84 | 86 | 85 |
| 1 day | 11 | 11 | 12 | 10 | 11 |
| $\geq$ 2 days | 4.0 | 4.0 | 4.4 | 3.8 | 3.9 |
| †Schizophrenia ED visits | | | | | |
| 0 | 98 | 97 | 98 | 98 | 98 |
| 1 visit | 1.6 | 1.8 | 1.3 | 1.4 | 1.6 |
| $\geq$ 2 visits | 0.7 | 0.9 | 0.4 | 0.5 | 0.6 |

†Measured in the 6 months before the index observation. All entries are percentages except for age. SSI = supplemental security income. ‡Includes affective disorders, anxiety disorders, obsessive-compulsive disorders, post-traumatic stress disorders, and other psychoses.

We first determined the number of classes, examining AIC and BIC statistics, fitting models with 1 through 5 classes. Next, we determined if unidimensionality held by comparing values of BIC for unidimensionality to BIC values for higher dimensions, fixing the number of classes at the value selected in the first step. The number of traits, $S$, was guided by the earlier factor analysis, with item allocation, $\mathcal{I}_s$, determined by the factor loadings. If an item loaded on more than one trait, we selected the trait with the highest loading. For identifiability, we set the discrimination parameter to 1 and the difficulty parameter to 0 for the first item in each trait.

## Quality Disparities

We used Equation (9) or (10) to estimate quality disparities depending on which latent trait distribution better fit the data.

### Validity

The validity of the latent trait(s) was assessed by (a) examining the mean difference in the latent trait (quality) between those having quarterly psychosocial visits (our held-out quality metric) and those who did not for the Normal latent trait model and the (b) probability of quarterly psychosocial visits across class membership groups for the latent class model. Positive differences (Normal latent trait) or larger probabilities of quarterly visits in higher classes (latent class model) support the validity of the model.

## Common Approach

We followed the common approach, computing an observed score for each enrollee according to Equation (1), estimating a linear regression model using the observed score as the response and the race and ethnicity variables as covariates. Scoring rules for $Y_j$ are provided in Supplement Table 1.

## Results

The final column in Table 2 describes the sample characteristics for the template sample of 24,000 individuals. Compared to the overall cohort (Column 1, Table 2), the template sample included more individuals from California; otherwise, all comorbid measures and health service use measures were no different from the full cohort. Covariates were balanced exactly across race and ethnicity groups (Supplement Table 2). Racial and ethnic quality differences persisted, e.g., quarterly ambulatory visits remained lowest for Black individuals (59% vs 67% Latinx) and for some items, the differences grew, e.g., 30-day follow-up of a schizophrenia emergency department visit (42% Black vs 53% White individuals; Supplement Table 3).

### *Exploratory Normal Latent Trait IRT Model*

In our initial analysis, the 1-factor IRT model (upper half, Table 3) had factor loadings ranging from 0 to 0.99 that were weighted towards acute care quality. In the two-factor model, factor 1 describes acute care quality while factor 2 describes pharmacological and ambulatory care quality. In the three-factor model, pharmacological care quality and ambulatory care quality are separated, although a few items (14: quarterly ambulatory care and 10: follow-up after an ED visit) do not load cleanly on a single factor. In the 2-factor and 3-factor models, the average pairwise correlation between item residuals was small. However, the discrimination parameters were very large for some quality indicators, and quantile-quantile plots indicated the non-normality of the estimated traits (Supplemental Figure 1).

We eliminated some items (1. long-acting injectable, 6. no polypharmacy, 16. no excessive mental health (MH) care, and 23. any psychotherapy) having low prevalence and poor item properties, and combined response options of some quality indicators. This resulted in 8 binary-valued indicators (bottom, Table 3). The distribution of these items are presented in Supplementary Table 4. The 2-factor solution continued to suggest acute care and ambulatory care quality traits. We fitted M2PL models (Equation 7) and while the RMSEA decreased, the latent trait distributions continued to demonstrate non-normality. Initial fitting suggested combining categories within items. This ultimately yielded binary-only items. Because the assumptions for the Normal latent trait model were not met, we abandoned the Normal latent trait model and did not estimate a structural regression model.

### *Exploratory Latent Class IRT Model*

Using the 8 binary items, we next fitted a latent class IRT model assuming 1 to 5 classes. The BIC suggested 3 classes and, while the AIC suggested 4 classes (Table 4), the difference between the 3 and 4 class model was negligible.

With 8 binary items, at most 2 dimensions were considered. We tested the 2-dimensional model in Equation 8 assuming an ambulatory care quality construct (quality indicators a, b, d, e, and h) and an acute care quality construct (quality indicators c, f, and g) against the unidimensional model. The constructs were identified by the earlier factor analysis. The BIC in Table 5 was lowest for 4-classes. Within a fixed number of classes, the likelihood ratio test rejected the assumption of unidimensionality.

The estimated class probabilities averaged over beneficiaries, and support points are presented in Table 6. Adjusting for race and ethnicity, the largest estimated class probability corresponds to the class receiving the highest care quality in the 3-class model (65% of the cohort) and only 9% of the cohort was characterized as belonging to the lowest quality class. The 4-class model included an additional higher class, removing a further 6% from the top 65% to form the highest quality group. The estimated support points reflect beneficiaries characterized by increasing quality, e.g., $\widehat{\xi}_2$ increases as $\widehat{\pi}_c$ increases. This is clear for both the 3 and 4-class models. However, the ambulatory care quality support points, $\widehat{\xi}_1$, do not increase smoothly as $\widehat{\pi}_c$ increases. In both the 3- and 4-class models, $\widehat{\xi}_1$ decreases as $\widehat{\pi}_1$ changes to $\widehat{\pi}_2$ although the size of the difference is smaller in the 4-class model.

Examination of the IRT parameter estimates suggest better item fit for the 3-class model and we use that model for inferences (the IRT parameter estimates for the 3-class

**Table 3.** Factor loadings, VARIMAX rotation using a matched sample of 24,000 beneficiaries.

| **No. of Factors** | **One** | **Two** | | **Three** | | |
|---|---|---|---|---|---|---|
| **Factor** | 1 | 1 | 2 | 1 | 2 | 3 |
| (2.) Long-acting injectable | 0.00 | 0.00 | **0.30** | 0.00 | 0.00 | **0.38** |
| (3.) Clozapine use | 0.08 | 0.00 | **0.24** | 0.02 | 0.00 | **0.33** |
| (4., 5.) Adherence | 0.12 | 0.07 | **0.25** | 0.11 | 0.05 | **0.29** |
| (6.) No polypharmacy | 0.00 | 0.00 | 0.00 | 0.00 | 0.00 | 0.00 |
| (7. - 9.) Follow-up after schizo. discharge | **0.88** | **0.98** | 0.09 | **0.99** | 0.00 | 0.01 |
| (10. - 12.) Follow-up after MH ED visit | 0.03 | 0.00 | 0.17 | 0.00 | **0.36** | **0.36** |
| (14.) Quarterly ambulatory MH visits | **0.24** | 0.09 | **0.99** | **0.23** | 0.00 | **0.86** |
| (15.) No excessive schizo. acute care | **0.92** | **0.91** | 0.10 | **0.83** | **0.56** | 0.00 |
| (16.) No excessive MH acute care | **0.87** | **0.87** | 0.08 | **0.75** | **0.65** | 0.00 |
| (19. , 20.) No readmission after MH discharge | **0.99** | **0.99** | 0.13 | **0.99** | 0.00 | 0.00 |
| (21.) Psychosocial services | 0.01 | 0.00 | **0.41** | 0.00 | 0.00 | **0.46** |
| (23.) Psychotherapy | 0.00 | 0.00 | **0.35** | 0.00 | 0.00 | 0.19 |
| Cumulative % Variation | 29 | 42 | | 47 | | |
| No. of Parameters | 30 | 46 | | 53 | | |
| AIC | 255243.5 | 258851.1* | | 245079.1 | | |
| RMSEA | 0.087 | 0.048 | | 0.067 | | |
| **No. of Factors** | **One** | **Two** | | | | |
| **Binary Transformed Indicators** | 1 | 1 | 2 | | | |
| (a.) Clozapine use | **0.26** | 0.00 | **0.35** | | | |
| (b.) Adherence $\geq$ 80 % | **0.32** | 0.12 | **0.38** | | | |
| (c.) Follow-up after schizo. discharge | **0.66** | **0.59** | **0.34** | | | |
| (d.) Follow-up after MH ED visit | 0.15 | 0.00 | **0.34** | | | |
| (e.) Quarterly ambulatory MH visits | **0.38** | 0.13 | **0.93** | | | |
| (f.) No excessive schizo. acute care | **0.90** | **0.88** | 0.01 | | | |
| (g.) No readmission after MH discharge | **0.94** | **1.00** | 0.00 | | | |
| (h.) Psychosocial services | 0.13 | 0.00 | **0.42** | | | |
| Cumulative % Variation | 31 | 46 | | | | |
| No. of Parameters | 16 | 23 | | | | |
| AIC | 133035.6 | 130245.8 | | | | |
| RMSEA | 0.026 | 0.015 | | | | |

Item 13 had no variation and was excluded. Item 22 was excluded and used as a validation measure. Loadings $\geq$| 20 | considered meaningful. 0.00 items that were estimated as 0 by the model. *The AIC is higher in the two factor model because some categories of items in the one factor models were combined.

**Table 4.** Fit of exploratory unidimensional latent class IRT models by the number of classes. Eight binary quality indicators.

| No. of Classes | Estimated Log Likelihood | No. of Parms. | AIC | BIC |
|---|---|---|---|---|
| 1 | -69209.50 | 8 | 138435.0 | 138499.7 |
| 2 | -66828.08 | 17 | 133690.4 | 133827.9 |
| 3 | -65675.85 | 19 | 131389.7 | **131543.3** |
| 4 | -65673.74 | 21 | **131389.5** | 131559.3 |
| 5 | -65675.66 | 23 | 131397.3 | 131583.3 |

**Table 5.** Unidimensional versus two-dimensional exploratory latent class IRT models.

| Model | Estimated Log Likelihood | No. of Parms. | AIC | BIC |
|---|---|---|---|---|
| 3-classes | | | | |
| Unidimensional | -65675.85 | 19 | 131389.7 | 131543.3 |
| Two-dimensional | -65505.40 | 20 | 131050.8 | 131212.5 |
| $-2$LogL | 340.9 | 1 dof | | |
| 4-classes | | | | |
| Unidimensional | -65673.74 | 21 | 131389.5 | 131559.3 |
| Two-dimensional | -65084.60 | 22 | 130227.2 | 130461.7 |
| $-2$LogL | 1178.3 | 1 dof | | |

model are presented in Supplementary Table 5). Quarterly ambulatory visits best discriminate ambulatory care quality among individuals having the largest $\widehat{a}_i = 3.49$ (se=0.62) with antipsychotic (AP) adherence of at least 80% having the next largest discrimination 1.06 (0.12). For acute care quality, both discrimination parameters were high: no excessive schizophrenia acute care, 4.16 (0.35), and no readmission following a mental health hospital discharge, 4.86 (0.53).

Figure 1 illustrates the estimated conditional response probabilities (y-axis) for each binary quality indicator. Class 3 had uniformly higher probabilities on the quality metrics relative to the other two classes. Relative to Class 1, individuals in Class 2 had higher probabilities on all acute care quality metrics but lower probabilities on all ambulatory care metrics.

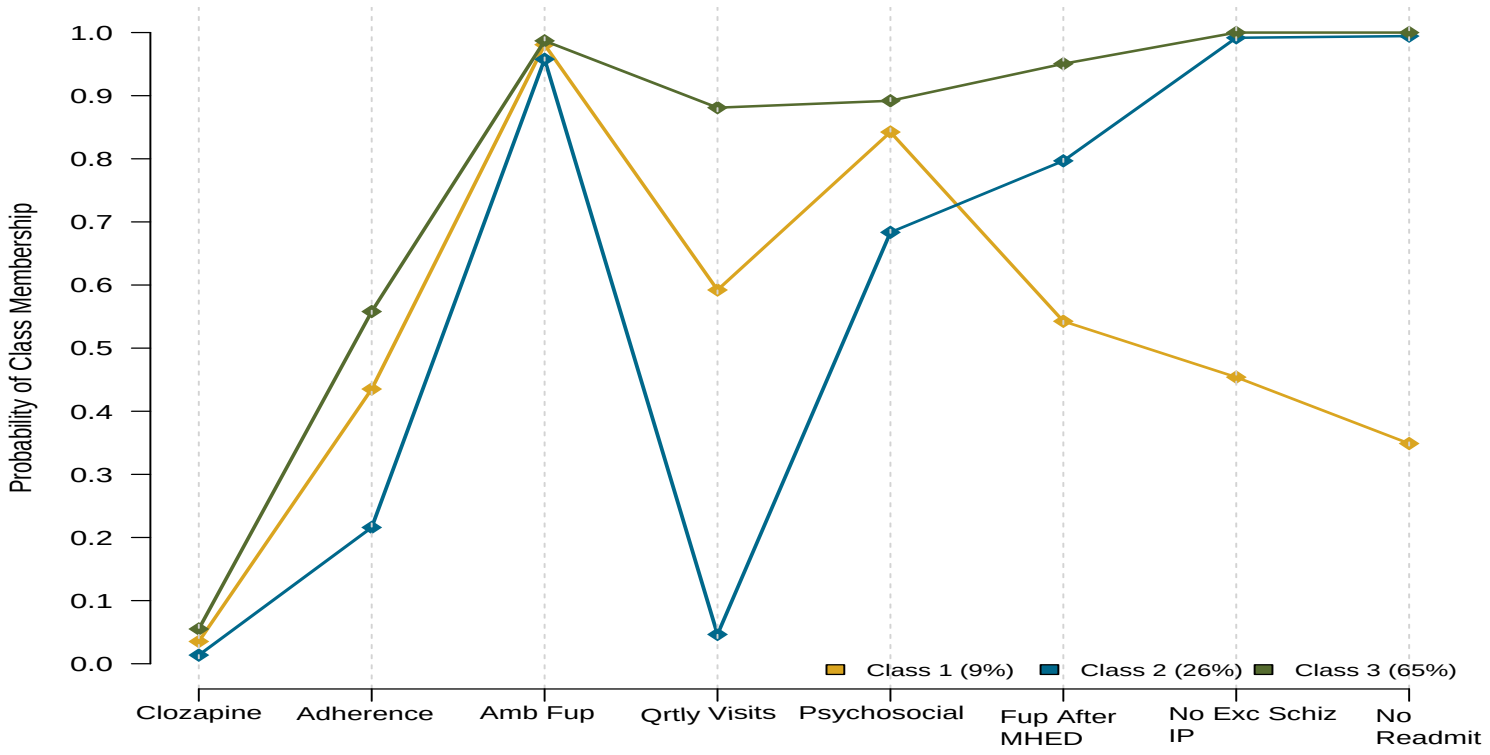

**Fig. 1.** Estimated healthcare quality profiles for 24,000 Medicaid beneficiaries with schizophrenia. Estimates use a 3-class 2-dimensional latent model adjusted for race and ethnicity using binary quality indicators listed in bottom half of Table 3.

**Table 6.** Estimated class probabilities and support points for exploratory 2-dimensional 3 and 4 class latent IRT models, adjusted for race and ethnicity group.

| Quality | | Ambulatory Care | | Acute Care | |
|---|---|---|---|---|---|
| **Class** | $\widehat{\pi}$ | $\widehat{\xi}_1$ | se($\widehat{\xi}_1$) | $\widehat{\xi}_2$ | se($\widehat{\xi}_2$) |
| 1 | 0.09 | -3.31 | 0.062 | 0.17 | 0.049 |
| 2 | 0.26 | -4.28 | 0.163 | 1.37 | 0.044 |
| 3 | 0.65 | -2.89 | 0.039 | 2.96 | 0.061 |
| 1 | 0.09 | -3.48 | 0.051 | 0.18 | 0.048 |
| 2 | 0.25 | -3.83 | 0.091 | 1.36 | 0.060 |
| 3 | 0.60 | -3.43 | 0.060 | 2.82 | 0.047 |
| 4 | 0.06 | -1.06 | 0.143 | 3.80 | 0.571 |

**Table 7.** Racial and ethnic quality disparity estimates using a matched sample of 24,000 individuals.

| | Black | | Latinx | |
|---|---|---|---|---|
| **Approach** | $\widehat{\gamma}_1$ | se($\widehat{\gamma}_1$) | $\widehat{\gamma}_2$ | se($\widehat{\gamma}_2$) |
| Observed Score Regression | -0.092 | 0.020 | 0.010 | 0.020 |
| | **Class 2 vs 1** | | **Class 3 vs 1** | |
| 3-Class Latent Model | $\widehat{\gamma}_1$ | se($\widehat{\gamma}_1$) | $\widehat{\gamma}_2$ | se($\widehat{\gamma}_2$) |
| White | 0.95 | 0.074 | 2.01 | 0.053 |
| Black vs White | 0.32 | 0.074 | -0.18 | 0.064 |
| Latinx vs White | 0.03 | 0.073 | 0.07 | 0.063 |

Observed score regression estimates are mean increases or decreases in quality relative to White beneficiaries on a linear scale. Latent class model estimates are changes in the log odds of the probability of a higher quality class versus Class 1 based on a two-dimensional model.

### *Validation*

While the percentage of individuals utilizing psychosocial services quarterly increased from 39% in Class 1 to 55% in Class 3, Class 2 had only 8%.

### *Estimated Quality Disparities*

The estimated racial and ethnic quality disparities are presented in Table 7 using the linear regression model and the 3-class 2-dimensional latent model using a multinomial logit model. The linear regression model indicates no differences in quality between Latinx and White beneficiaries. Black beneficiaries receive, on average, lower quality (0.092 standard deviations) of care. The regression coefficients for the latent class model are the log odds of the probability of class membership relative to the lowest quality class (9% of the sample). There is no difference between Latinx and White beneficiaries in quality class membership probabilities. Relative to White beneficiaries, Black beneficiaries are 1.4 times more likely than White beneficiaries to belong to class 2 versus 1 but only $\frac{4}{5}$ as likely to belong to the highest versus the lowest quality class.

Figure 2 displays $\widehat{\pi}_c$ for the sample and illustrates no difference in the distribution of class membership between White and Latinx beneficiaries. Relative to White and Latinx beneficiaries, fewer Black beneficiaries belong to the highest quality class. Figure 2 in the Supplement displays assignment to a single latent class, by race and ethnicity, using the maximum estimated probability (Equation 16).

## Conclusions

Item response theory (IRT) models can provide a unifying framework to estimate healthcare quality. This approach reflects the conceptual model of quality, accommodate features of the data-generating process, and permit joint modeling of the data useful for global testing. In our study, as in many healthcare quality studies, the quality metrics cover a broad set of services, such as pharmacological treatments, acute care services, and ambulatory services, making multidimensional IRT models realistic alternatives to commonly used approaches that assume a single quality construct. While approaches for estimation of unidimensional and multidimensional IRT models are widely available and frequently used in educational and psychological assessment settings, their application in the healthcare quality is infrequent. Most often, researchers seek a unidimensional solution given its simplicity and ease of computation. However, they may be inappropriate because quality constructs are likely multidimensional.

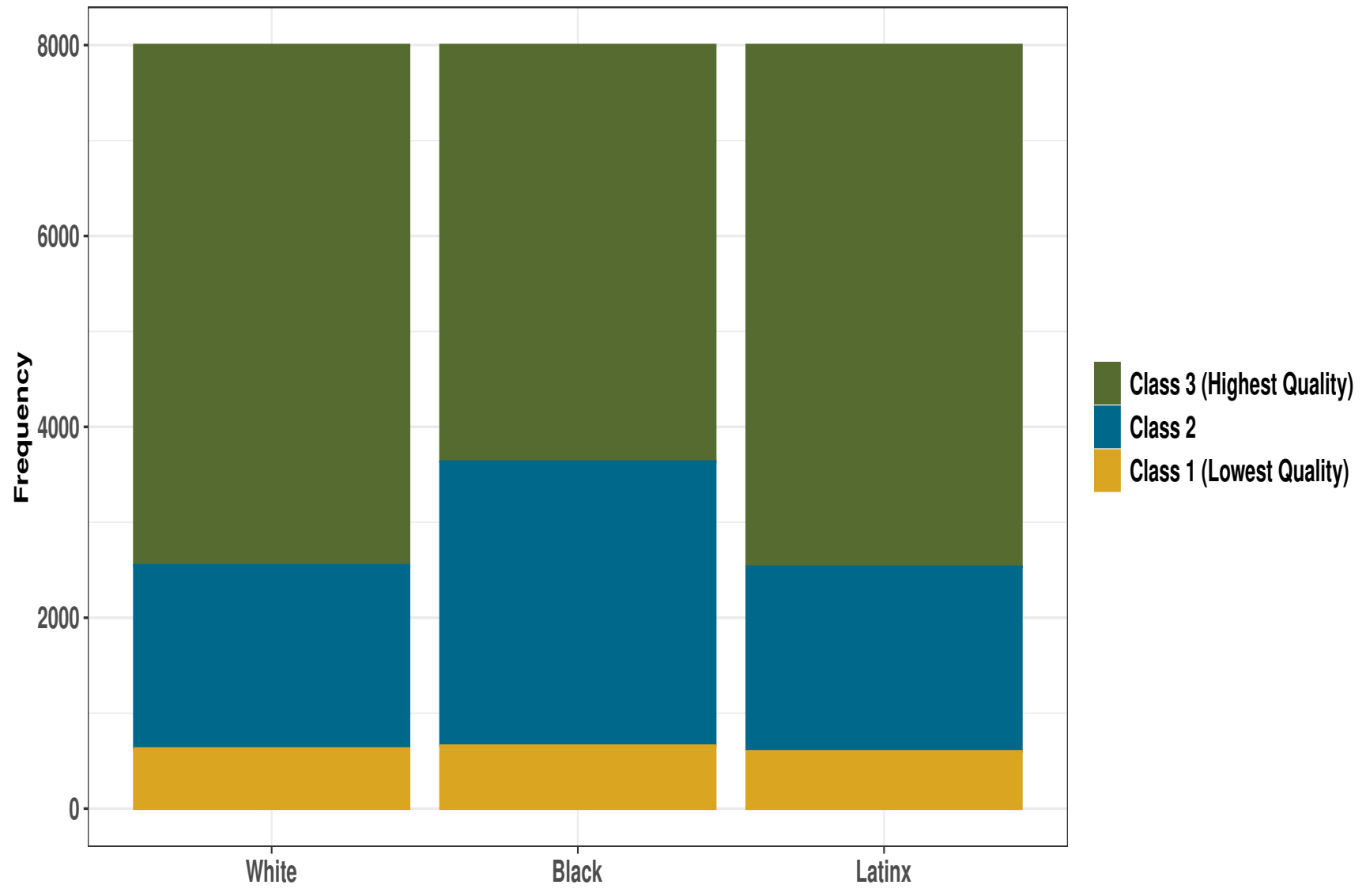

**Fig. 2.** Distribution of estimated latent class membership, by race and ethnicity status, for 24,000 Medicaid beneficiaries.

Parametric adjustments for differences in health status are challenging in general, but even more difficult when using structural regression models combined with IRT models. Identifiability issues are common and the number of covariates included at the trait level is limited. The use of template matching eliminates many parametric assumptions, simplifies any subsequent modeling required, and permits the inclusion of more confounders than possible via standard regression. Nonetheless, the IRT models are parametric as is the structural measurement model linking race and ethnicity to healthcare quality. Many researchers have proposed approaches to assess the sensitivity of results to unmeasured confounders in matched sets (see, in particular, Rosenbaum. (1987)), but we are unaware of their use in IRT models combined with structural regression.

In our study, unidimensionality did not fit the data nor did the Normal latent trait IRT model. Some metrics had response options such that a subset of them could be naturally ordered and others did not. For instance, when assessing if someone was followed up after a hospital discharge, options included no follow-up within 30 days, follow-up between 7 and 30 days, follow-up within 7 days, and no hospital discharge. While the first 3-options order naturally, having or not having a hospitalization could be indicative of good or bad quality of care. In our application, we ultimately combined most categories and arrived at binary items. The latent class IRT model provides a robust alternative to the Normal latent trait model, making no parametric assumptions about the latent trait. However, the estimates of the support points were not ordered for the ambulatory care quality construct, with the estimate of the location for class 2 imprecise. Estimating the items from the ambulatory care construct separately did not resolve this problem. The number of quality items was small in our study compared to health and educational assessments, preventing an accurate assessment of healthcare quality and quality disparities. Some quality metrics may be perceived as measures of a person's behavior, such as antipsychotic drug adherence, rather than measures of the quality of health services delivered. However, arguably the onus is on the healthcare system to educate patients and reduce any external barriers they may face to adhere to recommended care. Moreover, each measure used in the article is recommended by professional societies or other healthcare entities such as the National Committee for Quality Assurance. Not all quality metrics used in the study require risk adjustment. For instance, virtually all schizophrenia patients should be prescribed an antipsychotic drug. Given our goal of estimating disparities, we balanced health status characteristics across the race and ethnicity groups using information observed in billing claims data, but unmeasured confounding is always a risk. Finally, we did not have any patient preference information.

In summary, using an exploratory IRT measurement model combined with structural regression and template matching within a cohort of adults with schizophrenia, we found that healthcare quality is likely multidimensional and race and ethnicity disparities in quality likely exist. About 2/3 of the cohort was categorized into the highest-quality class, a quarter into the middle-quality class, and less than a tenth in the lowest class. While there were no differences between Latinx and White beneficiaries in the probability of belonging to higher-quality groups, Black beneficiaries were more likely than White beneficiaries to belong to the middle-quality class relative to the lowest-quality class but less likely to belong to the highest-quality class. Using the common approach of estimating a regression model using a score derived from the observed responses, we found no differences between Latinx and White beneficiaries, and a small difference between Black and White beneficiaries, with Black beneficiaries receiving poorer quality of care. These findings are clear for acute care quality – more work is needed for the ambulatory care quality construct.

## Acknowledgements

The authors thank Haley Abing, Harvard Medical School, for research program management and two Reviewers for helpful comments.

## Funding

This research was supported by the U.S. National Institute of Minority Health and Health Disparities Grant R01MD012428.

## Data Availability

The data used in this study are regulated by the Centers for Medicare and Medicaid Services (CMS) in the U.S. and are available for purchase under a data use agreement from CMS.

## References

Agniel D., Haviland A., Shekelle P., Scherling A., Damberg C.L. (2020). Distinguishing high-performing health systems using a composite of publicly reported measures of ambulatory care. *Annals of Internal Medicine*, 173: 791 – 798.

Anderson M.L., Nichol G., Dai D., Chan P.S., Thomas L., Al-Khatib S.M., Berg R.A., Bradley S.M., Peterson E.D. for the American Heart Association's Get With the Guidelines-Resuscitation Investigators. (2016). Association Between Hospital Process Composite Performance and Patient Outcomes After In-Hospital Cardiac Arrest Care. *Journal of the American Medical Assocation Cardiology*, 1(1):37–45.

Bacci S., Bartolucci F., Gnaldi M. (2014). A class of multidimentional latent class IRT models for ordinal polytomous item responses. *Communication in Statistics – Theory and Methods*, 43: 787–800.

Baker F. B. (1994). *Item Response Theory: Parameter Estimation Techniques*, New York, NY: Marcel Dekker Inc.

Bartolucci F. (2007). A class of multidimensional IRT models for testing for unidimensionality and clustering items. *Psychometrika*, 72 (2): 141–157.

Bartolucci F., Bacci S., Gnaldi M. (2014). MultiLCIRT: An R package for multicimensional latent class item repsonse models. *Computational Statistics and Data Analysis*, 71: 971–985.

Bartolucci F., Dardnanoi V., Peracchi F. (2016). Ranking scientific journals via latent class models for polytomous item response data. *Journal of the Royal Statistical Society, Series A*, 178 (4): 1025–1049.

Bennett M., Vielma J.P., Zubizarreta J.R. (2020). Building representative matched samples with multi-valued treatments in large observational studies. *Journal of Computational and Graphical Statistics*, 29(4):744 – 757.

Brown MW. (2001). An overivew of analytic rotation in exploratory factor analysis. *Multivariate Behavioral Research*, 36(1):111 – 150.

Busch AB, Frank RB, Sachs G. (2008). Bipolar-I depression outpatient treatment quality and costs in usual care practice. *Psychopharmacology Bulletin*, 41(2):24 –39.

Center for Medicaid and CHIP Services. (2023). *2023 Medicaid and CHIP Beneficiary Profile.* Baltimore, MD, Division of Quality and Health Outcomes, US Centers for Medicare & Medicaid Services.

Chalmers RP. (2012). mirt: A multidimensional item response theory package for the R environment. *Journal of Statistical Software*, 48(6): 1 – 29.

Cully JA, Zimmer M, Khan MM, Petersen LA. (2008). Quality of depression care and its impact on health service use and mortality among veterans. *Psychiatric Services*, 59(12): 1399 –1405.

Cummings J.R., Wen H., Ko M., Druss B.G. (2013). Geography and the Medicaid mental health care infrastructure: Implications for health care reform. *Journal of the American Medical Assocation Psychiatry*, 70(10):1084–1090.

Daniels MJ and Normand SL. (2006). Longitudinal profiling of health care units based on continuous and discrete patient outcomes. *Biostatistics*, 7(1):1–15.

Daskalakis C, Laird NM, Murphy JM. (2002). Regression analysis of multiple-source longitudinal outcomes: a "Stirling County" depression study. *American Journal of Epidemiology*, 155(1):88-94.

Druss B.G., Marcus S.C., Rosenheck R.A., Olfson M., Tanielian T., Pincus H.A. (2000). Understanding disability in mental and general medical conditions. *The American Journal of Psychiatry*, 157(9):1485 – 1491.

Embretson S.E. and Reise S.P. (2000). *Item Response Theory for Psychologists*, London, UK: Lawrence Erlbaum Associates.

Frank, R.G., and Glied, S.A. (2006). *Better but Note Well: Mental Health Policy in the US since 1950*, Baltimore, MD: Johns Hopkins Unversity Press.

Gerkens S, Maertens de Noordhout C, Lefèvre M, Levy M, Bouckaert N, Obyn C, Devos C, Scohy A, De Pauw R, Devleesschauwer B, Vlayen A, Yaras H, Janssens C, Meeus P. (2023). Performance of the Belgian health system: Revision of the conceptual framework and indicators set. *Health Services Research (HSR)*, Brussels, BE: Belgian Health Care Knowledge Centre (KCE). KCE Reports 370. DOI: 10.57598/R370C.

Goodman LA. (1974). Exploratory latent structure analysis using both identifiable and unidentifiable models. *Biometrika*, 61:215-–231.

He Y, Wolf RE, Normand ST. (2010). Assessing geographical variations in hospital processes of care using multilevel item response models. *Health Services Outcomes Research Methodology*, 10(3-4):111–133.

Hepner KA, Rowe M, Rost K, Hickey SC, Sherbourne CD, Ford DE, Meredith LS, Rubenstein LV. (2007). The effect of adherence to practice guidelines on depression outcomes. *Annals of Internal Medicine*, 147(5): 320 – 329.

Higashi T, Wenger NS, Adams JL, Fung C, Roland M, McGlynn EA, Reeves D, Asch SM, Kerr EA, Shekelle PG. (2007). Relationship between number of medical conditions and quality of care. *New England Journal of Medicine*, 24:2496–2504.

Horvitz-Lennon M., Donohue J.M., Domino M.E., and Normand S-L.T. (2009). Improving quality and diffusing best

practices: the case of schizophrenia, *Health Affairs*, 28(3): 701–712.

Institute of Medicine. (1990). *Medicare: A Strategy for Quality Assurance: Volume 1*. Washington, DC: National Academies Press.

Institute of Medicine. (2002). *Unequal Treatment: Confronting Racial and Ethnic Disparities in Health Care*. Washington, DC: National Academy Press.

Jank W. (2005). Quasi-Monte Carlo sampling to improve the efficiency of Monte Carlo EM, *Computational Statistics & Data Analysis*, 48(4):685–701.

Khaykin E, Eaton WW, Ford DE, Anthony CB, Daumit GL. (2010). Health insurance coverage among persons with schizophrenia in the United States. *Psychiatric Services*, 61(8):830 – 834.

Kilbourne AM, Farmer Teh C, Welsh D, Pincus HA, Lasky E, Perron B, and Bauer MS. (2010). Implementing composite quality metrics for bipolar disorder: towards a more comprehensive approach to quality measurement. *General Hospital Psychiatry*, 32(6): 636 – 643.

Kronick R., Gilmer, R., Dreyfus, T., and Lee, L. (2000). Improving health-based payment for Medicaid beneficiaries: CDPS., *Health Care Finance Review*, 21(3):29 – 64.

Landrum MB, Bronskill SE, Normand SL. (2000). Analytic methods for constructing cross-sectional profiles of health care providers. *Health Services and Outcomes Research Methodolog*, 1(1):23–48.

Landrum MB., Normand SL, Rosenheck RA. (2003). Selection of related multivariate means: monitoring psychiatric care in the department of veterans affairs. *Journal of the American Statistical Association*, 98:7–16.

Maydeu-Olivares A., Joe H. (2014). Assessing Approximate Fit in Categorical Data Analysis. *Multivariate Behavioral Research*, 49(4):305-328.

Minchin M, Roland M, Richardson J, Rowark S, Guthrie B. (2018). Quality of Care in the United Kingdom after Removal of Financial Incentives. *New England Journal of Medicine*, 379(10): 948 – 957.

Lex A, Gehlenborg N, Strobelt H, Vuillemot R, Pfister H. (2014). UpSet: Visualization of intersecting sets. *IEEE Transactions on Visualization and Computer Graphics*, 20(12): 1983 – 1992.

National Committee for Quality Assurance. (2010). *Medicaid Managed Care Quality Benchmarking Project: Final Report*. Washington, DC: National Committee for Quality Assurance.

Nolan, T. and Berwick, D.M. (2006). All-or-none measurement raises the bar on performance. *Journal of the American Medical Association*, 295(10):1168 – 70.

Normand S-L, Wolf RE, McNeil BJ. (2007). Discriminating hospital quality of care in the US. *Medical Decision Making*, 28:308–322.

Parks J, Svendsen D, Singer P, Forti M, eds. (2006). *Morbidity and mortality in people with serious mental illness*. Alexandria: National Association of State Mental Health Program Directors (NASMHPD) Medical Directors Council.

de Pouvourville G. (1997). Quality of Care Initiatives in the French Context. *International Journal for Quality in Health Care*, 9(3):163-–170, https://doi.org/10.1093/intqhc/9.3.163-a.

Ramalho A, Souza J, Castro P, Lobo M, Santos P, Freitas A. (2022). Portuguese Primary Healthcare and Prevention Quality Indicators for Diabetes Mellitus – A Data Envelopment Analysis. *The International Journal of Health Policy and Management*, 11(9):1725–1734.

Reckase MD. (2009). *Multidimensional Item Response Theory*. New York, NY: Statistics for Social and Behavioral Sciences, Springer.

Reise SP, Moore TM, Haviland MG. (2010). Bifactor models and rotations: exploring the extent to which multidimensional data yield univocal scale scores. *Journal of Personality Assessment*, 92(6):544–59.

Rosenabum, PR. (1987). Sensitivity analysis for certain permutation inferences in matched observational studies. *Biometrika*, 74, 13–26.

Substance Abuse and Mental Health Services Administration. (2019). *Behavioral Health Spending & Use Accounts 2006—2015*. HHS Pub. No. (SMA) 19-5095. Rockville, MD: Substance Abuse and Mental Health Services Administration.

Skrondal, A. and Rabe-Hesketh, S. (2004). *Generalized Latent Variable Modeling: Multilevel, Longitudinal, and Structural Equation Models*. New York, NY: Chapman & Hall/CRC.

Song Z, Rose S, Safran DG, Landon BE, Day MF, Chernew ME. (2014). Changes in health care spending and quality 4 years into global payment. *New England Journal of Medicine*, 371(18):1704–14.

Teixeira-Pinto A and Normand SL. (2008). Statistical methodology for classifying units on the basis of multiple-related measures. *Statistics in Medicine*, 27(9):1329–50.

Weinberger, D.R. and Harrison, P.J. (2011). *Schizophrenia*, 3rd Edition, Oxford, UK: Wiley-Blackwell.

Winkelman, D. and Damler, R. (2008). Risk adjustment in state Medicaid programs. *Health Watch*, 57: 14 – 17.

Yoon FB, Fitzmaurice GM, Lipsitz SR, Horton NJ, Laird NM, Normand SL. (2011). Alternative methods for testing treatment effects on the basis of multiple outcomes: simulation and case study. *Statistics in Medicine*, 30(16):1917-32.

# Supplementary Material: Estimating Racial and Ethnic Healthcare Quality Disparities Using Exploratory Item Response Theory and Latent Class Item Response Theory Models

Normand, Zelevinsky, and Horvitz-Lennon

The information presented in this Supplement describes results for the analysis of the quality of care information collected from 93,311 Medicaid enrollees residing in 5 U.S. states.

Table 1: *Scoring rules used to create observed score, $Y_j$. ED = emergency department; hosp. = hospitalization; NA = not applicable.*

| | |
|---|---|
| Any antipsychotic (AP) | 0=no; 1=yes |
| Long-acting injectible (LAI) | 0=no AP; 1=AP but not LAI; 2=LAI |
| Any clozapine | 0=no AP; 1=AP but not clozapine; 2=clozapine |
| AP adherence at least 80% | 0=no AP; 1=adherence $<$ 80%; 2=adherence $\geq$ 80% |
| AP adherence at least 50% | 0=no AP; 1=adherence $<$ 50%; 2=adherence $\geq$ 50% |
| No polypharmacy | 0=no AP; 1=polypharmacy; 2=no polypharmacy |
| Schizophrenia inpatient discharge | 0=discharge; 1=no discharge |
| Follow-up $\leq$ 7 days | NA=no discharge; 0= $>$ 7-days; 1= $\leq$ 7-days |
| Follow-up $\leq$ 30-days | NA=no schizo discharge; 0= $>$ 30-days; 1= $\leq$ 30-days |
| Schizophrenia ED visit | 0= $\geq$ 1 ED visit; 1=no ED visit |
| Follow-up $\leq$ 7 days | NA=no ED visit; 0= $>$ 7-days; 1= $\leq$ 7-days |
| Follow-up $\leq$ 30-days | NA=no ED visit; 0= $>$ 30-days; 1= $\leq$ 30-days |
| No schizoprenia hosp. | 0=hospitalized; 1=not hospitalized |
| No 30-day readmisison | NA=no hosp.; 0=readmitted $\leq$ 30-days; 1=not readmitted $\leq$ 30-days |
| No mental health (MH) hosp. | 0=hosp.; 1=not hospitalized |
| No 30-day readmission | NA=no hosp.; 0=readmitted $\leq$ 30-days; 1=not readmitted $\leq$ 30-days |
| Ambulatory MH care | 0=none; 1= some but $<$ quarterly; 2=quarterly |
| No excessive schizo acute care | 0=excessive care; 1=no excessive care |
| No excessive MH acute care | 0=excessive care; 1=no excessive care |
| Any psychosocial | 0=no psychosocial; 1=some but $<$ quarterly; 2=quarterly |
| Any psychotherapy | 0=none; 1= some |

Table 2: *Balance across race/ethnicity groups. Entries are column percentages.* †*State was not included because it is not a clinical need variable and potentially on the causal pathway.*

| **Characteristic** | **Template** | **Black** | **Latinx** | **White** |
|---|---|---|---|---|
| Number | 24,000 | 8,000 | 8,000 | 8,000 |
| Females | 46 | 46 | 46 | 46 |
| Age Quartile | | | | |
| $<$ 30 years | 25 | 25 | 25 | 25 |
| ([30, 43) years | 25 | 25 | 25 | 25 |
| [43, 52) years | 25 | 25 | 25 | 25 |
| $\geq$ 52 years | 25 | 25 | 25 | 25 |
| Index Year | | | | |
| 2010 | 53 | 53 | 53 | 53 |
| 2011 | 20 | 20 | 20 | 20 |
| 2012 | 14 | 14 | 14 | 14 |
| 2013 | 13 | 13 | 13 | 13 |
| Supplemental Security Income | 92 | 92 | 92 | 92 |
| Comorbidity | | | | |
| Cognitive | 3.9 | 3.9 | 3.9 | 3.9 |
| Psychiatric Conditions | 25 | 25 | 25 | 25 |
| Substance Use Disorder | 12 | 12 | 12 | 12 |
| Diabetes | 10 | 10 | 10 | 10 |
| Cardiometabolic | 43 | 43 | 43 | 43 |
| Other Medical | 21 | 21 | 21 | 21 |
| Schizophrenia Inpatient Days | | | | |
| 0 | 85 | 85 | 85 | 85 |
| 1 day | 11 | 11 | 11 | 11 |
| $\geq$ 2 days | 3.9 | 3.9 | 3.9 | 3.9 |
| Schizophrenia ED visits | | | | |
| 0 | 98 | 98 | 98 | 98 |
| 1 visit | 1.6 | 1.6 | 1.6 | 1.6 |
| $\geq$ 2 visits | 0.6 | 0.6 | 0.6 | 0.6 |
| †State of Residence | | | | |
| California | 62 | 37 | 90 | 58 |
| Georgia | 11 | 22 | 1.1 | 10 |
| Michigan | 14 | 22 | 2.2 | 19 |
| Mississippi | 4.6 | 10 | 0.1 | 3.6 |
| New Jersey | 8.4 | 9.3 | 6.6 | 9.3 |

Table 3: *Quality measures, overall and by race/ethnicity, for full cohort and matched sample. Entries are percentages unless specified otherwise.*

| | Full Sample | | | | Template | Matched Sample | | |
|---|---|---|---|---|---|---|---|---|
| **Metric** | **All** | **Black** | **Latinx** | **White** | **Sample** | **Black** | **Latinx** | **White** |
| Number of Persons | 93,311 | 41,297 | 14,301 | 37,713 | 24,000 | 8,000 | 8,000 | 8,000 |
| 1. Antipsychotic drug use | 90 | 90 | 93 | 90 | 91 | 89 | 93 | 90 |
| 2. Long-acting injectable | 15 | 17 | 14 | 13 | 14 | 17 | 13 | 13 |
| 3. Clozapine use | 4.6 | 2.5 | 4.7 | 7.0 | 4.7 | 2.5 | 4.4 | 7.1 |
| 4. Adherence $\geq$ 80% | 49 | 39 | 51 | 59 | 50 | 39 | 53 | 59 |
| 5. Adherence $\geq$ 50% | 72 | 64 | 76 | 80 | 74 | 64 | 78 | 80 |
| 6. No polypharmacy | 79 | 83 | 78 | 76 | 79 | 82 | 78 | 76 |
| 7. Schizophrenia discharge | 19 | 20 | 21 | 18 | 20 | 20 | 20 | 19 |
| 8. Follow-up care$\leq$ 7 days | 24 | 23 | 26 | 24 | 24 | 24 | 25 | 23 |
| 9. Follow-up care $\leq$ 30 days | 36 | 34 | 38 | 36 | 35 | 34 | 36 | 35 |
| 10. Schizophrenia ED visit | 3.6 | 4.4 | 2.3 | 3.1 | 3.1 | 4.1 | 2.2 | 2.9 |
| 11. Follow-up care $\leq$ 7 days | 37 | 35 | 38 | 39 | 37 | 32 | 39 | 44 |
| 12. Follow-up care $\leq$ 30 days | 47 | 44 | 51 | 51 | 46 | 42 | 45 | 53 |
| 13. Ambulatory MH care | 99 | 98 | 98 | 99 | 99 | 98 | 99 | 99 |
| 14. Qtly ambulatory visits | 63 | 58 | 66 | 66 | 64 | 59 | 67 | 65 |
| No excessive acute care for: | | | | | | | | |
| 15. Schizophrenia | 95 | 94 | 95 | 95 | 95 | 94 | 95 | 95 |
| 16. MH | 91 | 91 | 91 | 91 | 91 | 91 | 92 | 91 |
| 17. Schizophrenia hospital. | 20 | 20 | 22 | 19 | 20 | 21 | 20 | 20 |
| 18. No 30-day readmission | 77 | 77 | 76 | 77 | 77 | 77 | 76 | 78 |
| 19. MH hospitalization | 23 | 23 | 24 | 22 | 23 | 24 | 23 | 23 |
| 20. No 30-day readmission | 74 | 74 | 73 | 73 | 74 | 74 | 74 | 74 |
| 21. Psychosocial services | 84 | 85 | 83 | 83 | 83 | 85 | 82 | 83 |
| 22. Quarterly visits | 41 | 40 | 37 | 43 | 40 | 41 | 67 | 43 |
| 23. Psychotherapy | 28 | 36 | 12 | 26 | 25 | 37 | 13 | 26 |
| Mean number met (SD) | 10.0 (2.14) | 9.89 (2.21) | 10.0 (2.02) | 10.2 (2.08) | 9.6 (1.87) | 9.5 (1.97) | 9.7 (1.79) | 9.8 (1.84) |

Table 4: *Description and distribution of binary items.*

| | | **Percent** | |
|---|---|---|---|
| **Item** | **Comment** | **Full Cohort** | **Matched Sample** |
| (a.) Clozapine use | Any | 4.2 | 4.2 |
| (b.) Adherence | $\geq 80$ % | 44 | 46 |
| (c.) Follow-up after schizo. discharge | Any | 98 | 98 |
| (d.) Follow-up after MH ED visit | Any | 88 | 87 |
| (e.) Quarterly ambulatory visits | Quarterly | 62 | 64 |
| (f.) No excessive schizo. acute care | None | 95 | 95 |
| (g.) No readmit after MH discharge | None | 94 | 95 |
| (h.) Psychosocial services | Any | 88 | 83 |

Table 5: *Estimated exploratory latent class item response theory model parameters. Model parameters estimated from exploratory 3-class 2-dimensional model.*

| | **Ambulatory Care Quality** | | | |
|---|---|---|---|---|
| **Item** | $\widehat{a}_i$ | $se(a_i)$ | $\widehat{d}_i$ | $se(d_i)$ |
| Clozapine use | 1.00 | - | 0 | - |
| Adherence $\geq 80\%$ | 1.06 | 0.12 | -3.06 | 0.036 |
| Follow-up after schizo. discharge | 0.83 | 0.13 | -8.02 | 0.76 |
| Quarterly ambulatory visits | 3.49 | 0.62 | -3.42 | 0.082 |
| Psychosocial services | 0.93 | 0.12 | -5.11 | 0.26 |
| | **Acute Care Quality** | | | |
| **Item** | $\widehat{a}_i$ | $se(a_i)$ | $\widehat{d}_i$ | $se(d_i)$ |
| Follow-up after MH ED visit | 1.00 | - | 0 | - |
| No excessive schizo. acute care | 4.16 | 0.35 | 0.22 | 0.048 |
| No Readmit after MH discharge | 4.86 | 0.53 | 0.33 | 0.046 |

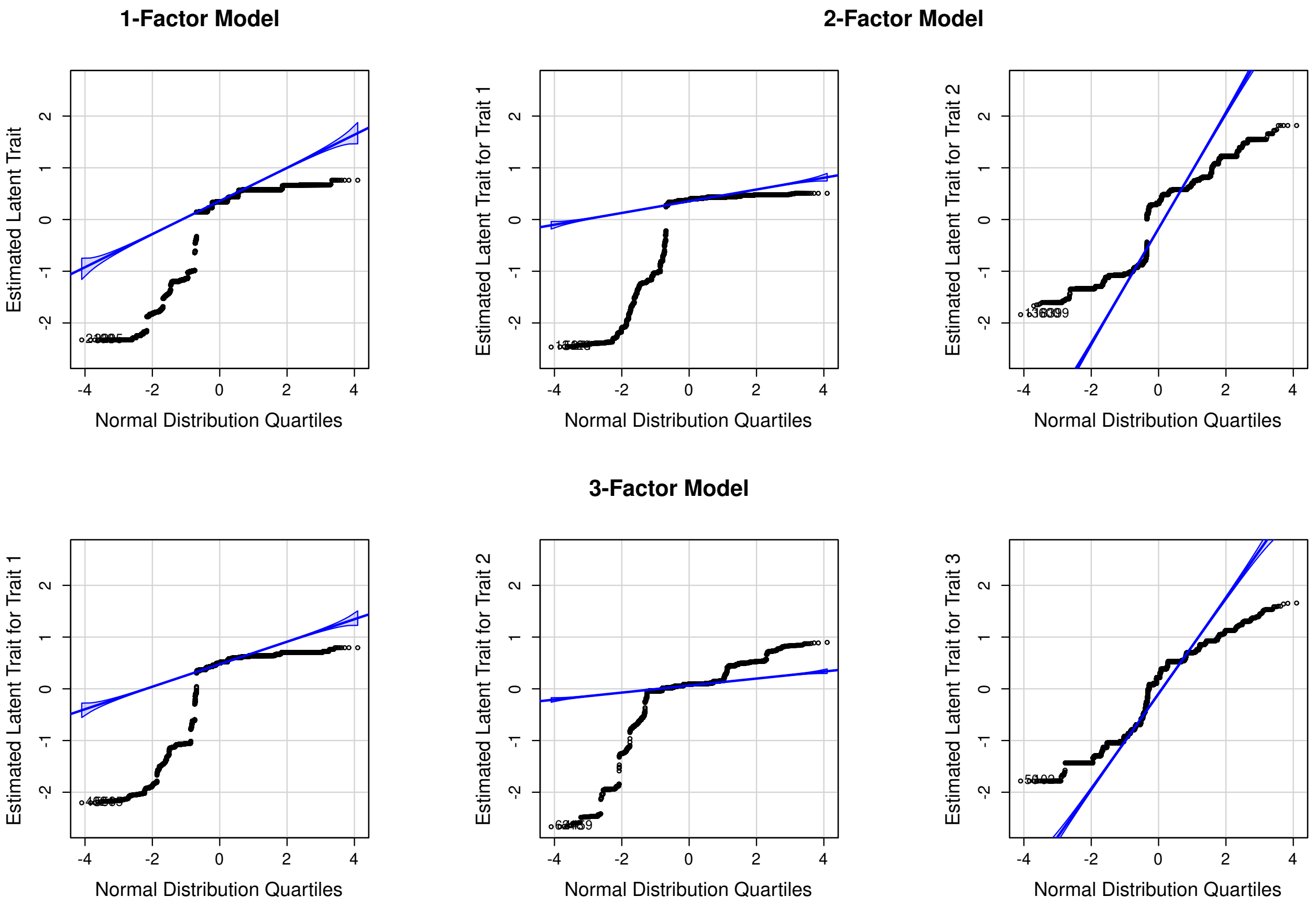

Figure 1: *Q-Q plots for exploratory Normal latent trait models. Estimated item response curves assuming Normal latent traits.*

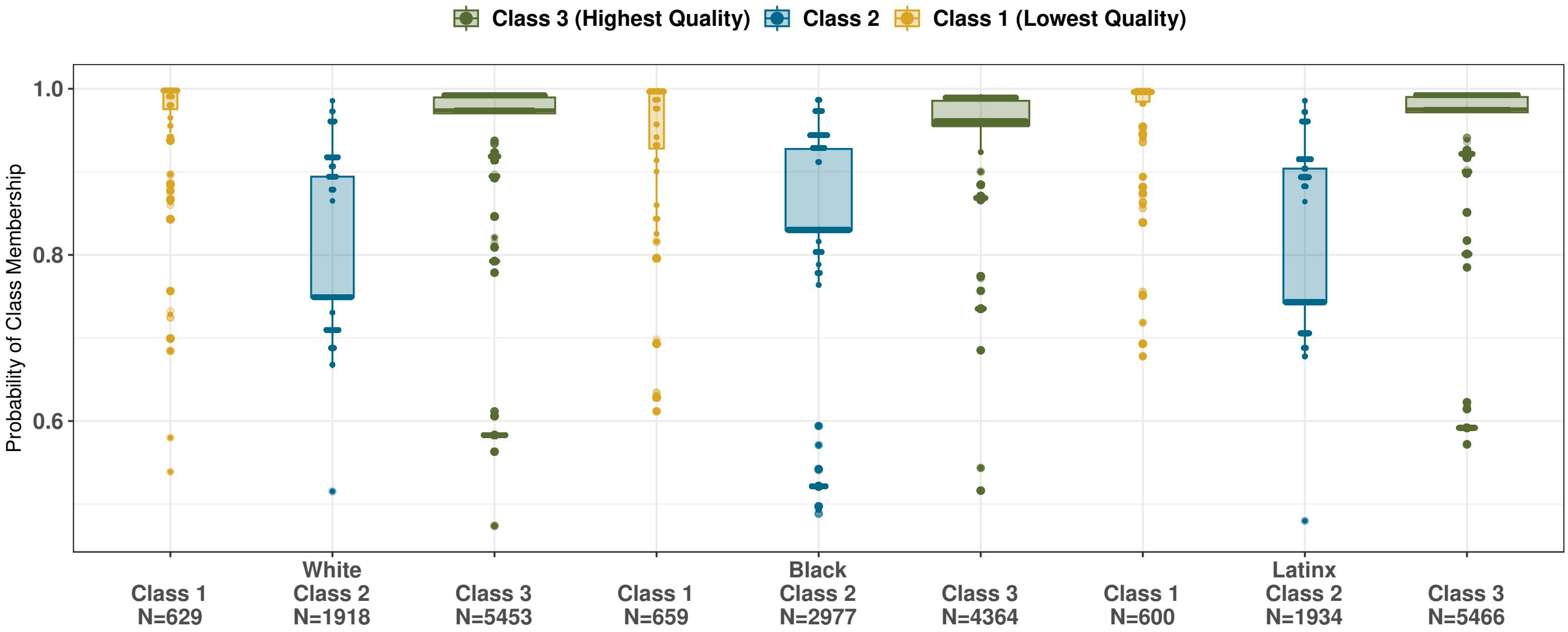

Figure 2: *Class assignment using maximum estimated probability.*